# On the Role of Dust Particles in Decoupling of Plasma from Magnetic Field in Laboratory and Space

L.I. Rudakov[1], A.V. Gretchikha[2], C.S. Liu[1], and G.M. Milikh[1]

**Abstract**

The effects of fast, non-diffusive penetration of the magnetic field into plasma with magnetized electrons and not magnetized ions have been actively studied in laboratories over the last 15 years. Applied to the space plasma, these effects can significantly change the picture of plasma dynamics. Charged dust particles are always present in the interstellar medium and can easily remain not magnetized, even when the plasma electrons and ions are strongly magnetized. In such a medium, the magnetic field can propagate with the super-Alfven velocity $cB/4\pi e \, |\nabla (1/z_d n_d)|$, where $z_d$ and $n_d$ are charge and density of the dust particles. This breaks the local "frozen-in" law for the magnetic field, which can significantly affect the gravitational collapse phenomena, the space plasma turbulence spectrum, magnetic dynamo and the rate of the Fermi acceleration of cosmic rays.

## 1. Introduction

The objective of this paper is to apply the recent achievements in understanding of the non-MHD dynamics of the magnetic field in plasma (acquired both in laboratory experiments, as well as in theory), to the interstellar phenomena. During the last 15 years, a significant interest was aroused by the physics of non-MHD penetration of a pulsed magnetic field into plasma. It is well known that magnetic field experiences skin effects in plasma. In collisional plasma, the classical skin depth is $(c^2t/4\pi\sigma)^{1/2}$, where $\sigma$ is the plasma conductivity. In collisionless plasma skin depth is $c/\omega_{pe}$, $\omega_{pe} = (4\pi n_e e^2/m)^{1/2}$. If the magnetic field is high enough, a compression shock wave moves out of the magnetic piston. In the shock wave, the magnetic field $B$ and density $n$ are compressed from the initial values $B_0$, $n_0$ up to the final values $B_m$, $n_m$. Ohmic dissipation occurs at the front of the shock wave, however the frozen-in low for the magnetic field is fulfilled, $B_m/B_o = n_m/n_o$. Such magneto-hydrodynamic (MHD) wave exists if $B_m/B_o < \gamma + 1/\gamma - 1$, $\gamma$ is the specific heat. Under higher values of the $B_m/B_o$ ratio an ordinary gas-dynamic compression wave is generated and the plasma ions must be heated [1].

---

[1] Department of Physics and Department of Astronomy, University of Maryland, College Park, MD 20742.
[2] Laboratory of Plasma Studies, Cornell University, Ithaca, NY 14853.

In laboratory experiment studying the dynamics of the magnetic field penetration into plasma, the installation represents a plasma-filled diode, with a current and magnetic field pulse generated at the edge of the diode. It was found that in low-density plasma, the magnetic signal propagates without noticeable plasma compression [2, 3]. The theory of this phenomenon is quite simple. Let us assume that the electrons are drifting in quasi-neutral plasma, while the ions are not magnetized and motionless. In this case, the dynamics of the magnetic field $B_y$ is described by the following equations

$$\frac{\partial B}{\partial t} + n_e \vec{V}_e \cdot \bar{\nabla}(\frac{B}{n_e}) = \frac{\partial}{\partial z} \frac{c^2}{4\pi\sigma} \frac{\partial B}{\partial z} \tag{1}$$

$$\vec{\nabla} \times \vec{B} = \frac{4\pi}{c} \vec{j} = -\frac{4\pi}{c} e n_e \vec{V}_e \tag{2}$$

Substituting $n_e V_e$ from (2), one obtains

$$\frac{\partial B}{\partial t} + \frac{c}{4\pi e} \frac{\partial}{\partial x}\left(\frac{1}{n}\right) B \frac{\partial B}{\partial z} = \frac{\partial}{\partial z} \frac{c^2}{4\pi\sigma} \frac{\partial B}{\partial z} \tag{3}$$

In a two-dimensional case, when $\nabla n$ along the flux of electrons $\vec{V}_e$ from cathode to anode exists, eq.(3) (so-called Burger's equation) has a nontrivial analytical solution. The magnetic field wave generates a steep front and converts into a shock wave $B = B_y(z - V_H t)$ with the front width $c^2/4\pi\sigma V_H$, where $V_H = \frac{cB}{8\pi e} \nabla_x \frac{1}{n}$. In order to neglect the acceleration of ions in the Hall electric field $E_z = -\frac{1}{c} V_{ex} B_y = \frac{1}{en_e} \frac{\partial}{\partial z} \frac{B_y^2}{8\pi}$, condition $V_H \gg V_A$ should be satisfied. This is the case if

$$\frac{c}{\omega_{pi}} \nabla_x \ell n\, n \gg 1, \quad \omega_{pi} = \left(\frac{4\pi e^2 n}{M}\right)^{1/2}. \tag{4}$$

The above results were obtained first in [4]. Later, they were generalized [5, 7, 8]. The ion motion in the shock wave was taken into consideration, and the criterion $c/\omega_{pi}\nabla \ell n\, n > 1$ was found for the shock wave formation. The shock wave also exists in a weekly collisional plasma; the thickness of its front is determined by scale size $c/\omega_{pe}$, while the ohmic dissipation is weak, i.e. behind the wave front $n_e T_e \ll B^2/8\pi$. The results of 2D simulations of the magnetic field penetration into plasma are presented in [5,6].

The physics of the described non-MHD shock wave is the following. The magnetic field is "frozen" into the electron fluid, according to eq. (1). In the traditional MHD, we assume that the electron and the ion velocities are close to each other, and therefore, the relative displacement of these two components (for the characteristic time scale) is small compared to the characteristic spatial scale $(\nabla \ell n\, n)^{-1}$. In our case, the electron displacement is finite. Electrons transfer the frozen-in magnetic flux and the magnetic energy. They preserve the "frozen-in" condition ($B/n = const$) along the wave front, moving towards the increasing density $n$ in the magnetic field, which increases in time. However, there is no local "frozen-in" law for ions in such a wave, in contrast to a conventional MHD wave. In the laboratory plasmas ($n = 10^{12} - 10^{14}$ cm$^{-3}$), the spatial scale, for which Hall MHD (HMHD) is applicable, is 3..30 cm. Such diodes are used as plasma switches for rapid current switching on the load (the wave front is sharpening!). During last decade physics of the HMHD and Plasma Opening Switches (POS) was discussed at numerous physics meetings, some reviews were published [9].

## 2. Dynamics of the magnetic field in space dusty plasma

In the laboratory, magnetosphere or space plasmas, the HMHD phenomena are usually studied for the scales of the order of $c/\omega_{pe}$, where they are important for the description of stochastization and reconnection of the magnetic field lines [10]. Recently, in paper [11], eq. (3) was applied for the description of the magnetic fields dynamics in neutron stars, where the ions form a motionless lattice. For the space plasma, the ion collisionless scale $c/\omega_{pi}$ ~$10^2 – 10^4$ km is too small to affect the global dynamics of the interstellar plasma. However, a significant amount of charged dust particles always exists in the space plasma. The charges are mostly negative and can be easily estimated based on the theory of a Langmuir probe with floating potential, $z_d \cong - 3\times10^7 aT(eV)$. For $a = 1\mu m$ (this corresponds to the particle mass $m \cong 4 \times 10^{-12} g$) and $T = 1\ eV$, the charge is $z_d = - 3 \times 10^3$ electron charges. In space, dust particles can play the role of heavy unmagnetized ions, while the plasma electrons and ions are magnetized for the astrophysical temporal and spatial scales. Let us consider such 3-component quasi-neutral ($n_i - n_e = -z_d n_d$) plasma. We assume that the dust particles are non-magnetized and moving under electrical and gravitational fields

$$\frac{dV_d}{dt} = \frac{z_d e}{M_d}\vec{E} - \nabla\psi, \qquad \Delta\psi = 4\pi G M_d n_d, \qquad (5)$$

while electrons and ions are drifting with the velocities

$$\vec{V}_e = c\frac{\vec{E}\times\vec{B}}{B^2}, \quad \vec{V}_i = c\frac{\vec{E}\times\vec{B}}{B^2} + \frac{v_i}{\omega_{ci}}c\frac{\vec{E}}{B} \qquad (6)$$

$$\vec{\nabla}\times\vec{B} = \frac{4\pi e}{c}\left[c\frac{\vec{E}\times\vec{B}}{B^2}(n_i - n_e) + \frac{v_i}{\omega_{ci}}c\frac{\vec{E}}{B}n_i + z_d n_d \vec{V}_d\right] =$$
$$= \frac{4\pi e}{c}\left[-z_d n_d c\frac{\vec{E}\times\vec{B}}{B^2} + \frac{v_i}{\omega_{ci}}c\frac{\vec{E}}{B}n_i + z_d n_d \vec{V}_d\right] \qquad (7)$$

Here $z_d$, $n_d$, are the charge and the density of the dust particles, $v_i$ is the rate of scattering of the plasma ions by neutrals and dust particles. In this section, we will consider the cylindrical geometry $B = B_\varphi(r,z)$, $n_d = n_d(r)$ as shown in Fig. 1 and assume that dust particles are motionless, $V_d = 0$. Eq.(7) can be resolved relative to $E$. We assumed here that the motion takes place almost across the magnetic field $B$, and that $\vec{E}\cdot\vec{B} = 0$:

$$\frac{4\pi e n_i \vec{E}}{B} = \frac{-z_d n_d / n_i (\vec{\nabla}\times\vec{B}\times\vec{B}/B) + v_i/\omega_{ci} \vec{\nabla}\times\vec{B}}{(z_d n_d/n_i)^2 + (v_i/\omega_{ci})^2}, \qquad (8)$$

Taking *curl* of the eq. (8), one obtains the equation similar to the frozen-in equation (3)

$$\frac{\partial B_\varphi}{\partial t} + \frac{cr^2}{8\pi e}\vec{\nabla}B^2 \times \vec{\nabla}\frac{r^{-2} z_d n_d/n_i}{(z_d n_d/n_i)^2 + (v_i/\omega_{ci})^2} = -(\vec{\nabla}\times\eta_i\vec{\nabla}\times\vec{B})_\varphi \qquad (9)$$

$$\eta_i = \frac{v_i M c^2}{4\pi e^2 n_i}\frac{1}{(z_d n_d/n_i)^2 + (v_i/\omega_{ci})^2} \qquad (10)$$

It is important to mention that equations (9), (10) are different from the equations describing the evolution of magnetic field in the frame related to the mass center of the charged particles which is normally used in astrophysical plasma studies [12-15].

A new term in the left part of eq. (9) has appeared. The magnetic diffusion coefficient $\eta_i$ due to the ambipolar drift, first introduced by Spitzer [12] also differs by the factor of $(v_i/\omega_{ci})^2/((n_d z_d/n_i)^2 + (v_i/\omega_{ci})^2)$. Since the dust particles are always present in the space plasma, this coefficient can be less than unity. As shown in [15], in hydrogen molecular clouds, the value $z_d n_d/n_i$ changes from a small value up to unity when $n_H$ spans from $10^{-3}$ cm$^{-3}$ to $10^{12}$ cm$^{-3}$ during the gravitational collapse. The value $v_i/\omega_{ci}$ remains small, and the ambipolar drift $\eta_i \frac{\partial B}{\partial z}$ can be less than the one calculated using the Spitzer formula.

The reduction of ambipolar drift has been taken into account in [15], however the second term in the left part of eq. (9) was missed. This was caused by the specific geometry of the magnetic field considered in [15] (the magnetic field was directed along $z$, while the current was flowing along y (in our case $\varphi$). Based on the symmetry considerations, the authors of [15] assumed that the electric drift of both ions and electrons preserves the local "frozen-in" law $B/n = const$. The geometry is used in this paper is more suitable for description of the disk-like gravitational collapse, see Fig.1, with the magnetic field $B = B_\varphi$, and the electrons and the ions drifting along the plasma density gradient, and carrying the frozen-in magnetic field along the radius. Therefore, the local "frozen-in" law is broken in our case. Similar to our discussion in the Introduction, criterion for neglecting the dust particles motion in the Hall shock wave of magnetic field $B_\varphi = B_\varphi(z - V_{Hd}t)$

$$V_{Hd} = \frac{r^2 c B_\varphi}{8\pi e} \nabla_r \frac{r^{-2} z_d n_d}{(z_d n_d)^2 + (v_i n_i / \omega_{ci})^2} \qquad (11)$$

is $V_{Hd} > B/(4\pi n_d M_d)^{1/2}$. The scale $L_d$ for which this approximation is still valid:

$$L_d = [\nabla_r \ln(r^2 z_d n_d)]^{-1} < \left(\frac{M_d c^2}{4\pi e^2 z_d^2 n_d}\right)^{1/2}; \quad M_d n_d >> M n_i \qquad (12)$$

The second criterion arises if we remember that the ion inertial drift current $d(c E/B\omega_{ci})/dt$ was neglected in eq.(7) compared to the term $z_d n_d cE/B$. Also dust particle motion was neglected in eq.(8). It is right if $dV_d/dt >> \omega_{cd} V_d$, $\omega_{cd} = z_d eB/cM_d$. Finally, the criterion is $M_d n_d >> M n_i$.

The parameters of Dust Clouds listed in review [16] are: $n_d = 10^{-7}$ $cm^{-3}$, $n_i = 10^{-3}$ $cm^{-3}$, $a = 1\mu$, $M_d = 4 \times 10^{-12}$ g, $B = 3\mu G$. For $T_{eV} = 1$ $eV$, $z_d = -3 \times 10^3$, $L_d = 3\times 10^{13}$ $cm \cong 2$ $AU$. On the limit of applicability of Hall MHD, $n_d \geq M n_i/M_d$, $L_d < 2\times 10^{18}$ $cm \cong 1$ $pc$.

It is well accepted that energy equipartition occurs in the Universe, which means that the magnetic, turbulent kinetic, thermal and gravitational energy densities are roughly equal. This can prevent the development of gravitational collapse. Indeed, the energy of the magnetic field frozen in the matter increases faster than the gravitational energy and hinders the collapse [12-16]. The above analysis shows that such hindering could be absent at scales smaller than $L_D$.

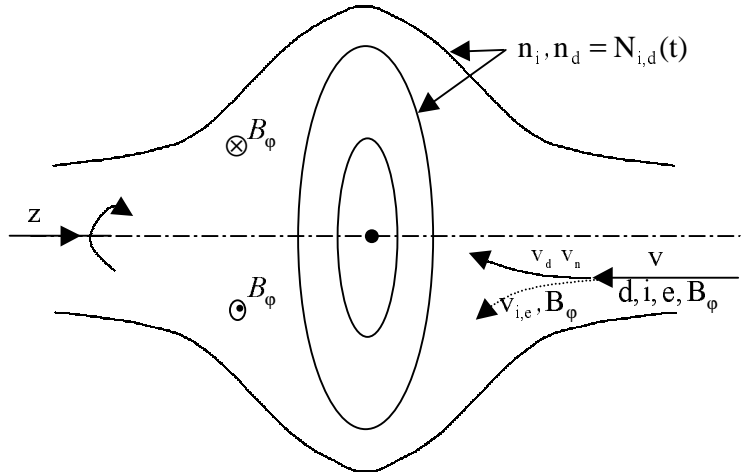

**Fig 1. Sketch of Gravitational Collapse**

The plasma and frozen-in magnetic field fluxes during gravitational collapse are qualitatively shown on Fig. 1. The ions and electrons are flowing outwardly along the lines $B/n = const$, while the dust particles and hydrogen molecules are collapsing on the disk along $z$.

## 3. Oscillations and instabilities in an inhomogenious dusty plasma

The oscillations and instabilities of dusty plasma have been extensively studied [17-19]. In contrast to paper [18] where the electrostatic oscillations were considered we will focus on the electromagnetic oscillations of dust clouds with the parameters mentioned above, as in [16]. Let us consider the case of plasma which is inhomogenious along x, the magnetic field $B_o$ directed along *y*, and the wavenumber *k* of perturbations directed along *z*. The magnetic field is strong enough to prevent gravitational collapse of the dusty plasma cloud. It takes place if gravitational potential $\psi$ is less than $B^2/8\pi M_d n_d$. According to eq.(5) it is fulfilled if the size of cloud *L* is less than $L_{JB} = B/4\pi M_d n_d G^{1/2}$. A plasma slab inhomogenious along *x* is not necessary in equilibrium – it could be in 1D nonlinear motion. But we are looking for 2D oscillations with phase velocities much higher than $B_0/(4\pi M_d n_d)^{1/2}$. Let the perturbations of all values be proportional to $exp(-i\omega t + ikz)$, $\omega >> dlnB_0/dt$, $k >> (\nabla lnn_d, \nabla lnB_0)$, $kB_0 >> \vec{k}\cdot\vec{B}_0$, $L_{JB}(\nabla lnn_d) > 1$. In accordance with the physics discussed above we assume that the oscillating electrons and ions are magnetized and cold. The plasma is quasi-neutral ($n_i - n_e = -z_d n_d$). We assume that the ion friction with neutrals and dust particles is small and neglect the last term in (6). In the "dust Hall MHD" case, $\omega >> kV_{Ad}$, $\omega_{cd}$, which occurs when inequalities (12) are fulfilled, the dust current evaluated by eq.(5) and continuity equation is $i\omega z_d^2 e^2 n_d E/M_d(\omega^2 + \omega_J^2)$ and less than the drift current $z_d e n_d E/B$. For this reason we can approximate the Hall electric field by (8), where $v_i/\omega_{ci} << zn_d/n_i$. In the same time we take into account the dust motion along $k$ direction under the Hall electric field $-\nabla B^2/8\pi z_d e n_d$. It results in changes in the dust density and current: $\delta n_d/n_d = kV_d/\omega = k^2 B_0 * \delta B/4\pi n_d M_d(\omega^2 + \omega_J^2)$, $(\delta j_d)_z = - z_d e \delta n_d E_{0x}/cB_0 + i\omega z_d^2 e^2 n_d \delta E_z/M_d(\omega^2 + \omega_J^2) = \delta n_d/4\pi n_d \, dB_0/dx - \omega k z_d e B_0 \delta B/4\pi M_d(\omega^2 + \omega_J^2)$. After linearizing eqs. (5-7) we can combine in the left part of eq.(7) terms which are functions of $\delta B$, then calculate $E$ and take *curl* of $E$, as we did when we derived eq.(9). Then we obtain:

$$\frac{\omega - kV_{Hd} + ik^2\eta_i}{\omega - kV_{Bd}} = \frac{k^2 V_{Ad}^2}{\omega^2 + \omega_J^2}, \qquad (13)$$

$$V_{Hd} = -\frac{cB_0}{4\pi z e}\frac{\nabla_x n_d}{n_d^2}, V_{Bd} = -\frac{c}{4\pi z e}\frac{\nabla_x B_0}{n_d}, V_{Ad} = \frac{B_0}{(4\pi M_d n_d)^{1/2}},$$

where $\omega_J = (4\pi G n_d M_d)^{1/2}$ is the Jeans frequency. For the case a homogenious plasma and $\omega >> \omega_J$ dispersion relation (13) describes the dust magnetic sound waves in the frequency range $(\omega_{ci}\omega_{cd})^{1/2} >> \omega >> \omega_{cd}$.

If we neglect the motion of dust particles in (13), we will arrive at the drift mode of the electromagnetic oscillations corresponding to linearized eq. (9) when $z_d n_d/n_i >> v/\omega_{ci}$.

$$\omega = \frac{kcB_o}{4\pi e}\nabla_x \frac{1}{z_d n_d} - ik^2\eta_i \qquad (14)$$

The drift mode (14) takes place at $\omega_{ci} >> \omega >> \omega_{cd}$, $kV_{Ad}$ or:

$$\frac{4\pi e^2 n_i}{M_i c^2}\frac{z_d n_d}{n_i} > \frac{k}{L_d} > \frac{4\pi z_d^2 e^2 n_d}{M_d c^2} \qquad (15)$$

If we take into account dust motion while assuming the plasma is homogeneous we get similarly to [16] that short scale Jeans instability is suppressed by magnetic pressure:

$$\omega^2 = k^2 V_{Ad}^2 - \omega_J^2 \qquad (16)$$

If $\nabla B_o / B_o \ll \nabla n_o / n_o$ equation (13) has an unstable solution for shorter scales:

$$\left(\frac{L_d^2 4\pi z_d^2 e^2 n_d}{M_d c^2}\right)^{1/2} < kL_{JB} < 2\left(\frac{M_d c^2}{L_d^2 4\pi z_d^2 e^2 n_d}\right)^{1/2}. \qquad (17)$$

Opposite to the conclusion of paper [16] that magnetic pressure is an obstacle for Jeans instability on the scale $k^{-1} < V_{Ad}/\omega_J$, we got for inhomogenious dusty plasma the possibility of gravitational collapse on smaller scales.

Now we consider Hall MHD instability of current carrying plasma, when $\omega \gg \omega_J$.

$$\frac{\omega - kV_{Hd} + ik^2\eta_i}{\omega - kV_{Bd}} = \frac{k^2 V_{Ad}^2}{\omega^2}. \qquad (18)$$

This equation has unstable solution if $V_{Bd} > V_{Ad}$ (see p.288 in [9]). It corresponds to the criteria:

$$L_{Bd} = (\nabla_x \ln B_0)^{-1} < \left(\frac{M_d c^2}{4\pi e^2 z_d^2 n_d}\right)^{1/2}, \frac{\nabla \ln B_0}{\nabla \ln n_d} > 1. \qquad (19)$$

Solution of eq. (18) is easy to find if $\nabla B_o / B_o \gg \nabla n_o / n_o$.

$$\mathrm{Im}\,\omega = 3^{-1/2} k \, (V_{Ad}^2 V_{Bd})^{1/3} = 3^{-1/2} kV_{Ad}\left(\frac{M_d c^2}{4\pi e^2 z_d^2 n_d L_{Bd}^2}\right)^{1/6} \qquad (20)$$

Instability develops when $k^2\eta_i > \omega$. This fast instability was found in 2-D simulation of plasma dynamics in frame of Hall MHD [5,6] and in theory in [9]. Unstable Hall wave was also found in simulations [5] when the shock was initiated by magnetic piston applied to the boundary of homogeneous plasma.

## 4. Conclusions

Dust particles are always present in the solar and interstellar plasma. At the scales

$$\left(\frac{M_i c^2}{4\pi e^2 z_d n_d}\right)^{1/2} < L_d < \left(\frac{M_d c^2}{4\pi e^2 z_d^2 n_d}\right)^{1/2},$$

their presence can significantly change the dynamics of the magnetic field compared to that described by the MHD equations if $M_d n_d > M n_i$. The decoupling of the magnetic field and plasma dynamics caused by the dust component has to be considered at astrophysical scales. Presently, this effect is neglected by the existing theories. On the limit of applicability of the Hall MHD (12) and for typical value $n_i = 10^{-3} \, cm^{-3}$ this spatial interval covers a wide range from $10^{10}$ cm to $10^{18}$ cm. On this scales Jeans gravitational instability can developed easily, see (17). The presence of dust particles can reduce the escape of the magnetic field out of molecular clouds due to the ambipolar ion drift (10), but can also lead to the dynamic escape of the magnetic field (9). The existence of the magnetic field shock waves (9) with the propagation velocity (11) and the "Hall potential drop" across the front $B^2/8\pi z_d n_d e$ (may be in $M_d/M \, z_d$ times bigger than for usual magnetic sound wave) can significantly increase the rate of the Fermi acceleration of cosmic rays. Fast instability of nonlinear motion of the Hall plasma opens a new opportunities for transfer of turbulent energy into small scales, see (20). The escape of magnetic field can result in some restrictions on magnetic dynamo process.

We have described the new phenomena using very simplified model. In reality dust particles are distributed over the size. The mass density of dust is defined probably by the particles of the biggest size while the charge is determined by the smaller size particles.

Thus the problem should be considered by introducing a distribution function of dust particles over the size. The above analysis was made for simple magnetic field geometry. The sensitivity of this phenomenon to the field geometry has to be investigated. We neglected the thermal pressure of plasma components since the computer simulations [5,6] along with the theory [8] had shown that plasma heating due to magnetic field penetration does not change the discussed physics.

After the paper has been written we discussed it with Professor P. Shukla who turned our attention to the fact that the convective term exists in his work [19], which is devoted to derivation of the equations and to the analysis of electromagnetic oscillations in dusty plasma with nonuniform density. However the importance of this term for the formation of the magnetic field shock wave first discussed in [4] was not mentioned. The gravitational effects have not been considered in [19] and the plasma was assumed to be in equilibrium, while the currents were absent. Thus the effect of reduction of the threshold of gravitational instability, as well as instability of the nonlinear motion when $\nabla B_o / B_o > \nabla n_o / n_o$ (see eqs. 17-21) were missing in [19].

**Acknowledgments.**

The authors are grateful to Prof. R. Z. Sagdeev and Prof. K. Papadopoulos for their interest in this work and for their support during the course of this research. We are also grateful to Prof. P. Shukla for useful discussions.